\documentclass[twocolumn,prl,showpacs,superscriptaddress]{revtex4}
\usepackage{graphicx,amssymb,amsmath}
\begin{document}
\title{Electron spin resonance signal of Luttinger liquids and single-wall carbon nanotubes}
\author{B. D\'ora}
\email{dora@pks.mpg.de}
\affiliation{Max-Planck-Institut f\"ur Physik Komplexer Systeme, N\"othnitzer Str. 38, 01187 Dresden,
Germany}
\author{M. Gul\'{a}csi}
\affiliation{Max-Planck-Institut f\"ur Physik Komplexer Systeme, N\"othnitzer Str. 38, 01187 Dresden,
Germany}
\author{J. Koltai}
\affiliation{Department of Biological Physics, E\"{o}tv\"{o}s University, P\'{a}zm\'{a}ny
P\'{e}ter s\'{e}t\'{a}ny 1/A,
1117 Budapest, Hungary}
\author{V. Z\'{o}lyomi}
\affiliation{Research Institute for Solid State Physics and Optics of the Hungarian
Academy of Sciences, P. O. B. 49, H-1525, Budapest, Hungary}
\author{J. K\"{u}rti}
\affiliation{Department of Biological Physics, E\"{o}tv\"{o}s University, P\'{a}zm\'{a}ny P\'{e}ter s\'{e}t\'{a}ny 1/A,
1117 Budapest, Hungary}
\author{F. Simon}
\affiliation{Budapest University of Technology and Economics,
Institute of Physics and Condensed Matter Research Group of the
Hungarian Academy of Sciences, H-1521, Budapest P.O.Box 91, Hungary}

\date{\today}

\begin{abstract}
A comprehensive theory of electron spin resonance (ESR) for a Luttinger liquid (LL) state of correlated metals is
presented. The ESR measurables such as the
signal intensity and the line-width are calculated in the framework of Luttinger liquid theory with broken spin 
rotational symmetry as a function of magnetic field
and
temperature.
We obtain a significant temperature dependent homogeneous line-broadening which is related to the spin symmetry 
breaking and the electron-electron interaction. The result crosses over
smoothly to the ESR of itinerant electrons in the non-interacting limit.
These findings explain the absence of the long-sought ESR signal of itinerant electrons in single-wall carbon 
nanotubes when considering realistic experimental conditions.

\end{abstract}


\maketitle

The experimental and theoretical studies of strong correlation effects are in the forefront of condensed matter research.
Low-dimensional carbonaceous systems, fullerenes, carbon nanotubes (CNTs), and graphene exhibit a rich variety of such phenomena including superconductivity in
alkali doped fullerenes \cite{GunnRMP}, quantized transport in SWCNTs,
and massless Dirac quasi-particles showing a half integer
quantum Hall-effect in graphene even at room temperature \cite{NovoselovNature}.
A compelling correlated state of
one-dimensional systems is the Luttinger liquid (LL)
state. There is now abundant evidence from both theoretical \cite{eggergog,kane,balents,yoshioka,krotov}
and experimental \cite{bockrath,BachtoldPRL2004,ishii,rauf} side that the low energy 
properties of CNTs with a single shell, the single-wall carbon
nanotubes (SWCNTs) can be described with the LL state.
As a result, SWCNTs are regarded as a model system of the LL state,
which could be exploited to further test the theories and novel experimental methods.

Electron spin resonance (ESR) is a well established and powerful method to characterize correlated states of itinerant electrons.
It helped to resolve e.g. the singlet nature of superconductivity in elemental metals
\cite{VierSchultz}, the magnetically ordered spin-density state in low-dimensional organic metals \cite{TorrancePRL1982} and
in alkali doped fullerides \cite{JanossyPRL1997}. In three-dimensional
metals, the ESR signal intensity is proportional to the Pauli spin-susceptibility,
the ESR line-width and $g$-factor are determined by the mixing of spin up and down states due
to spin-orbit (SO) coupling in the conduction band. These ESR measurables are affected when
correlations are present and thus their study holds information about the nature of the correlated
state.

This motivated a decade long quest to find the ESR signal of
itinerant electrons in SWCNTs and to characterize its properties in the
framework of the expected correlations \cite{PetitPRB1997,NemesPRB2000,SalvetatPRB2005}. Detection of ESR in SWCNTs is also vital for applications as it enables to determine the spin-lattice relaxation time, $T_1$, which determines the usability for spintronics \cite{FabianRMP}.
However, to our knowledge no conclusive evidence for this observation has been reported.
An often cited argument for this anomalous absence of the ESR signal is the large
heterogeneity of the system, the lack of crystallinity, and the presence of magnetic catalyst
particles \cite{PetitPRB1997,NemesPRB2000,SalvetatPRB2005}.
However, ESR signal of conduction electrons has been indeed observed for
electron doped SWCNTs \cite{NemesPRB2000}, which are known to be Fermi liquids rather than the pristine SWCNTs \cite{rauf}. Thus the above properties of SWCNTs
should hinder the observation of the ESR signal also for the Fermi liquid state, which is clearly not the case.
As a result, we suggest that the LL state inherently prohibits the observation of ESR of the itinerant electrons,
calling for a realistic description of such experiments. A recent experiment by Kuemmeth \textit{et al.} \cite{kuemmeth} shed new light on the spin degree of freedom of SWCNTs.
It was shown that SO coupling and correspondingly the lifting of the spin rotational
invariance is unexpectedly large. As we show below, this results in a uniquely large homogeneous
broadening of the ESR line which explains the absence of an intrinsic ESR signal of SWCNTs.
So far, the theory of ESR in the SWCNTs was limited to SU(2) symmetric models \cite{martino,martino1}.

Here, we study the ESR signal in a Luttinger liquid with broken spin rotational symmetry.
While at low temperatures the characteristic non-integer power laws characterize the response, the high temperature behavior
crosses over to the standard Lorentzians, whose width, in contrast to the Fermi liquid picture \cite{Slichterbook}, 
is determined by the Luttinger liquid parameters.
We show that this explains the absence of itinerant electron spin resonance in this
system by combining DFT calculations of the spin-susceptibility on metallic SWCNTs with a critical evaluation of the experimental conditions.

To describe a metallic SWCNT, we apply an effective low-energy theory. We neglect the "flavour" index coming from the two $K$ points \cite{martino1} since we are interested in the spin properties only.
The standard Luttinger liquid Hamiltonian is expressed as a sum of independent spin and charge excitations as
\begin{equation}
H=\sum_{\nu=c,s}\frac{\hbar v_\nu}{2}
\int \textmd{d}x\left(K_\nu\Pi^2_\nu+\frac{1}{K_\nu}(\partial_x\phi_\nu)^2\right),
\label{hamilton}
\end{equation}
where $K_\nu$'s are the Luttinger liquid parameters, $\nu=c,s$ denotes the charge and spin sector,
respectively, $\Pi_\nu$ and $\phi_\nu$ are canonically conjugate fields with velocity $v_\nu$. The Luttinger liquid parameter
in the spin sector, $K_s=1$ for SU(2) symmetric models as these preserve the spin rotational
symmetry. However, the presence of spin-orbit and magnetic dipole-dipole interaction between the conduction
electrons \cite{spinanis,nersesyan} produces spin dependent interactions and breaks the spin rotational symmetry, leading
to $K_s\neq 1$. In addition, these processes are also responsible for the $g$-factor anisotropy.

The original fermionic field operators are expressed in terms of the bosons as
\begin{gather}
\Psi_{r\sigma}(x)=\frac{\eta_{r\sigma}}{\sqrt{2\pi\alpha}}\times \nonumber \\
\times\exp
\left(i\sqrt \frac\pi
2\left(r\phi_c(x)+r\sigma\phi_s(x)+\Theta_c(x)+\sigma\Theta_s(x)\right)\right),
\end{gather}
which are needed to express the spin operators in the bosonic language, $\eta_{r\sigma}$ is the Klein factor,
$\Theta_{\nu}(x)=-\int_{-\infty}^x\textmd{d}y\Pi_\nu(y)$, $r$=R/L=$\pm$ denotes the chirality of the
electrons, and $\sigma=\pm$ is the electron spin.

The ESR experiments are performed in a longitudinal static magnetic field, $B$,
applying a transversal perturbing microwave radiation with a magnetic component, $B_{\perp}$. For the ESR
description, the above Hamiltonian is completed with the Zeeman term:

\begin{equation}
H_{Z}=-g \mu_{\textmd{B}} {B}\int \textmd{d}x\partial_x\phi_s(x),
\end{equation}

The ESR signal intensity is given by the absorbed microwave power that is \cite{Slichterbook}:

\begin{equation}
I(\omega)=\frac{B_{\perp}^2\omega}{2 \mu_0}\chi^{\prime\prime}_{\perp}(q=0,\omega)V,
\end{equation}
where $\mu_0$ is the permeability of the vacuum, $\chi^{\prime\prime}_{\perp}$ is the imaginary part of the
retarded spin-susceptibility for the transversal direction, and $V$ is the sample volume.
The spin operators required to calculate $\chi^{\prime\prime}_{\perp}$ are $S^\pm(x)=\sum_{r,r'}
\exp(i(r'-r)k_F)\Psi_{r\pm}^+(x)\Psi_{r'\mp}(x)$. Since ESR measures the $q=0$ response, only the $r=r'$
terms contribute, the others contain fast oscillating terms $\sim \exp(\pm2ik_F)$ and average to zero.

The Zeeman term in Abelian bosonization is the simplest when the longitudinal magnetic field points in the spin 
quantization axis (the $z$-axis).
For a different field orientation, the Zeeman term becomes more complicated but it can be rotated along the $z$ direction,
at the expense of changing the Luttinger liquid parameters $K_{\nu}$ \cite{spinanis}.
The external magnetic field further lowers the SU(2) symmetry in addition to the spin-orbit
and dipole-dipole interactions, resulting in a further renormalization of $K_s$.

From now on, we set $\hbar=k_B=g\mu_{\text{B}}=1$ and they will be reinserted whenever necessary. The retarded spin-susceptibility is built up from correlators of the type \cite{schulz}
\begin{gather}
\langle S^+(x,t)S^-(0,0)\rangle=c_\perp^2\left(\frac{\pi T\alpha/v_s}{\sinh[\pi
T(x/v_s-t+i\alpha)]}\right)^{2+\gamma}\times\nonumber\\
\times \left(\frac{\pi
T\alpha/v_s}{\sinh[\pi T(x/v_s+t-i\alpha)]}\right)^{\gamma}\exp\left(\frac{ibx}{v_s}\right),
\end{gather}
where $b=K_s{B}$, $c_\perp$ is determined by the short distance behavior and cannot be obtained by the methods used here. Here we introduced the
$\gamma=(K_s+1/K_s-2)/2\approx (\delta K_s)^2/2$ parameter, where $K_s=1+\delta K_s$, and $\delta K_s$ encodes information about
spin symmetry breaking processes. Upon Fourier transformation, we obtain the retarded spin-susceptibility.
From a simple scaling analysis, we can conjecture the behaviour of the retarded spin-susceptibility as $\chi(q=0,\omega)_{\perp}\sim \textmd{max}[{B},\omega,T]^{2\gamma}$
which is confirmed later by a careful investigation in Eqs. \eqref{esrt0} and \eqref{esrapprox}.

Putting all this together, we find for the ESR intensity:
\begin{gather}
I(\omega)=-A\sin(\pi\gamma)\omega
\left(\frac{2\pi\alpha
T}{v_s}\right)^{2\gamma}\times\nonumber \\
\times\textmd{Im}\left[F(2+\gamma,k_1)F(\gamma,k_2)+F(2+\gamma,k_2)F(\gamma,k_1)\right],
\label{esrint}
\end{gather}
where
\begin{gather}
k_{1,2}=\frac{\omega\mp b}{2\pi T}\\
F(x,y)=B\left(\frac{x-iy}{2},1-x\right),
\end{gather}
where $B(x,y)=\Gamma(x)\Gamma(y)/\Gamma(x+y)$ is Euler's beta function, $\Gamma(x)$ is Euler's gamma function, $A$ is a constant, whose value
is determined further below. In the $\gamma=0$ limit, SU(2) spin symmetry is conserved by the Hamiltonian and the ESR resonance becomes completely
sharp, located at $\pm B$ as $\sim {B}^2 \delta(\omega\pm {B})$.

The influence of interactions is most clearly seen at $T=0$, when the ESR
signal is completely asymmetric around $\omega=\pm b$, and cannot be
approximated by Lorentzians:
\begin{gather}
I(\omega)=A\left(\frac{\alpha }{2 v_s}\right)^{2\gamma}\sin^2(\pi\gamma)
\frac{\Gamma^2(1-\gamma)}{\gamma(1+\gamma)}\times\nonumber\\
\times
\frac{2|\omega|(\omega^2+b^2)}{(\omega^2-b^2)^{1-\gamma}}\Theta(|\omega|-b).
\label{esrt0}
\end{gather}
The ESR intensity vanishes below a threshold set by the magnetic field, and
falls off in a power law fashion, depending on the explicit value of $K_s$.

\begin{figure}[h!]
{\includegraphics[width=7cm,height=7cm]{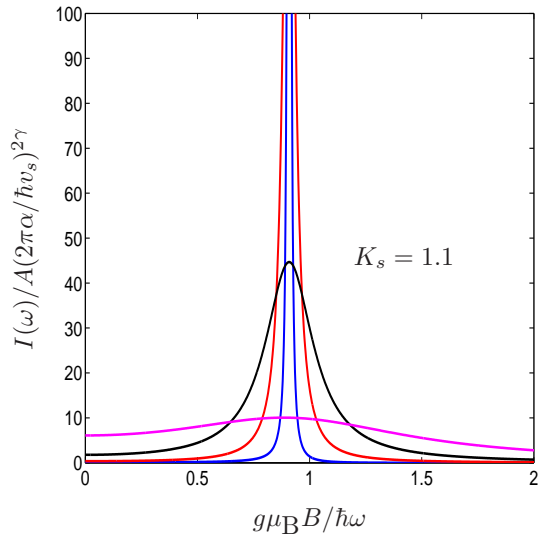}}

\caption{(Color online) The ESR \textit{signal} intensity,
Eq. \eqref{esrint} is shown as a
function of the magnetic field for $K_s=1.1$ (corresponding to $\gamma=0.0045$), $k_BT/\hbar\omega=0.1$ (blue),
1 (red), 5 (black) and 25 (magenta). The resonance occurs at
$g\mu_{\text{B}}{B}/\hbar\omega=1/K_s\approx 0.91$
for small temperatures.
The approximate formula, Eq. \eqref{esrapprox} cannot be distinguished from the
exact result on the scale of the figure. As the temperature increases, the
ESR
line broadens significantly and the tail of the Lorentzian centered at
negative
field extends to the positive field window as well, resulting in one single
broadened
peak at $B=0$.
\label{esr0p01}}
\end{figure}

However, the sharp threshold disappears with increasing temperature and the spectrum broadens. 
In the limit of $T\gg \omega,{B}$ and $\gamma\ll 1$, which is relevant for realistic experiments,
the intensity can be approximated by (upon reinserting original units)
\begin{gather}
I(\omega)=A 2\pi(\hbar\omega)^2\times\nonumber\\
\times\left[\frac{\eta}{(\hbar\omega-K_sg\mu_{\text{B}}{B})^2+\eta^2}+
\frac{\eta}{(\hbar\omega+K_sg\mu_{\text{B}}{B})^2+\eta^2}\right]
\label{esrapprox}
\end{gather}
where $\eta=2\gamma\pi k_B T$. This expression works well outside of its range of validity and it consists of two Lorentzians, centered
around $\pm K_sg\mu_{\text{B}}{B}$, characterized by a width of $\eta$. Hence, the interaction ($\gamma$) together 
with the
temperature determines the width of the resonance and shifts the resonance center as well,
 as is seen in Fig. \ref{esr0p01}.

This expression allows us to make contact with the conventional Fermi liquid case.
In that
case, $K_s=1$ (together with $\gamma\rightarrow 0$), and the ESR intensity reduces to
\begin{equation}
I(\omega)=A2\pi^2 (g
\mu_{\textmd{B}}{B})^2\left[\delta(\hbar\omega-g\mu_{\text{B}}{B})+
\delta(\hbar\omega+g\mu_{\text{B}}{B})\right].
\end{equation}
Thus the integrated ESR intensity reads as $\int \textmd{d}\omega I(\omega)/\omega=A4\pi^2 g \mu_{\textmd{B}}{B}$.
In a Fermi liquid, this is expressed in terms of the static
spin-susceptibility \cite{Slichterbook}, $\chi_0$,
as $\int \textmd{d}\omega I(\omega)/\omega=\chi_0 B_{\perp}^2 V\pi g \mu_{\textmd{B}}{B}/2 \mu_0 \hbar$.
This fixes the so far unknown numerical prefactor as $A=\chi_0 B_{\perp}^2 V/{2 \mu_0 \hbar \pi}$. In summary, the ESR signal of a Luttinger liquid with broken spin rotational symmetry i) is significantly broadened due to the interaction and spin symmetry breaking and ii) has a signal intensity which matches that of the non-interacting state.

These results are similar to those found for the 1D antiferromagnetic Heisenberg model \cite{oshikawa}, whose low
energy theory is identical to the spin sector of a Luttinger liquid, Eq. \eqref{hamilton}. The  ESR
line-width also scales with $T$
at low temperatures.
However, the spin in the Heisenberg model, when
represented in terms of fermionic variables via the Jordan-Wigner transformation, usually contains non-local string
operators \cite{giamarchi} and acquires a different scaling dimension than
the spin of itinerant electrons.
The exchange anisotropy, causing the broadening of the ESR signal, shares common origin with the $g$-factor anisotropy
in terms of spin-orbit coupling, scaling with $(\Delta g/g)^2$.

The spin orbit coupling in SWCNTs was found to be unexpectedly large, around
1~meV for a nanotube with diameter of 1~nm, resulting in a $g$-factor enhancement $g=2.14$ in a few electron carbon
nanotube quantum dot \cite{kuemmeth}. In the presence of many electrons,
the interplay of interactions, low dimensionality and spin-orbit coupling determines
the strongly correlated ground state and it can further enhance spin symmetry breaking.
 $K_s\sim 1.3$ for quantum wires \cite{gritsev} like InAs, which is another possible realization of Luttinger liquids. These materials possess a spin orbit coupling of the same order of
magnitude than SWCNTs but they have a smaller Fermi velocity.
Following a similar line of reasoning for SWCNTs, we take a conservative estimate of $K_s$ to be around
1.1.

In the following, we discuss the relevance of the results on the absence of an ESR signal in SWCNTs.
As we showed above, the ESR signal intensity of a Luttinger liquid crosses over smoothly for the 
non-correlated case to the static susceptibility that is the
Pauli susceptibility of metallic SWCNTs \cite{ashcroft}: $\chi_0=\mu_0 g^2/4 \mu_{\textmd{B}}^2  D(E_{\textmd{F}})$, 
where $D(\epsilon_{\textmd{F}})$ is
the density of states (DOS) at the Fermi energy. To have an accurate value for the DOS in a realistic sample, we 
performed density functional theory calculations with the Vienna {\it ab
initio} Simulation Package \cite{KresseG_1996_2} within the local
density approximation for metallic
nanotubes. The projector augmented-wave
method was used with a plane-wave cutoff energy of 400~eV. The DOS
was obtained with a Green's function approach from the band
structure that was calculated with a large $k$-point sampling. We considered the (9,9),
(15,6), (10,10), (18,0), and (11,11) SWCNTs in order of increasing diameter. Here $(n,m)$ 
denotes the lattice vectors on the graphene basis along which a cut-out stripe is folded up 
to represent a SWCNT \cite{DresselhausTubes}. These tubes are within the Gaussian diameter 
distribution of a usual SWCNT sample with a mean diameter of 1.4~nm and a variance of 0.1~nm. 
Calculating the DOS for chiral SWCNTs with a large number of atoms in the elementary cell is 
prohibitively long. Therefore, we confirmed by nearest-neighbor tight binding calculations 
on \textit{all} the metallic SWCNTs in the above diameter distribution that the DOS depends 
very weakly on the chirality, thus the above SWCNTs chiralities are indeed representative for 
the ensemble of the metallic tubes.

We obtain that such a tube ensemble has an effective DOS of $D(E_{\textmd{F}})=4.6\cdot 10^{-3}$~states/eV/atom 
by averaging the DOS for the above SWCNTs and taking into account that only one third of the tubes are metallic 
for this diameter range \cite{DresselhausTubes}. This is a very low DOS which results from the one-dimensionality 
of the tubes and from the fact that the majority of the tubes are non-metallic. It is 50 times smaller than 
in K$_3$C$_{60}$ ($D(E_{\textmd{F}}) \approx 0.3$~states/eV/atom \cite{GunnRMP}) and is comparable to the 
well known low DOS of pristine graphite ($D(E_{\textmd{F}}) \approx 5\cdot 10^{-3}$~states/eV/atom
\cite{Dresselhaus_AP_Review}). With the above DOS, we obtain that a typical 2~mg SWCNT sample 
gives a practically detectable signal-to-noise ratio of $S/N=10$ for a spectrum measured for 1000 seconds 
provided the ESR line is not broader than 110~mT. To obtain this value, we considered that the state-of-the-art 
ESR spectrometers give a $S/N=1$ for $10^{10}$ S=1/2 spins at 300~K provided the ESR line-width is 0.1~mT and 
each spectra points (typically 1000) are measured for 1~sec. We also took into account that the $S/N$ drops with 
the square of the line-width for broadening beyond 1~mT.

The above calculated homogeneous broadening of the ESR line of a Luttinger liquid is 
$2 \pi \gamma k_BT/g \mu_{\text{B}}$ in units of the magnetic field. Thus at 4~K, which is the
lowest available temperature for most ESR spectrometers, one has a broadening of $\gamma \cdot 18.7$~Tesla. 
This, together with the above detectability criterion gives
an upper limit of $\gamma=6\cdot 10^{-3}$ for the detection of the ESR 
signal \footnote{This is an overestimate of the limit as many other factors such as more limited
microwave penetration into the sample makes the ESR experiment less sensitive and thus reduce this 
limit, which further justifies our argument.}. Clearly, the above conservative estimate 
of $\gamma=4.5 \cdot 10^{-3}$ based on the the $K_s=1.1$ value is close to this limit, 
which explains why careful studies have not yet yielded a conclusive ESR signal of itinerant 
electrons is SWCNTs. This argument can be also turned around: the fact that no ESR signal of 
the itinerant electron has been observed in the SWCNTs means that the line is broadened beyond 
observability, which means that the real $\gamma$ is larger than $6 \cdot 10^{-3}$ putting also $K_s>1.1$.

We finally comment on the future viability of this observation. Clearly, ESR spectrometers operating to sub 
Kelvin temperatures are required. Observation of linearly temperature dependent ESR line-width would be an 
unambiguous evidence for the observation of the ESR signal of itinerant electrons in the Luttinger liquid 
state. We note that such a temperature dependence is fairly unusual as ESR line-width in metals normally 
tends to a residual value similar to the resistivity.

In summary, we extended the theory of electron spin resonance in a Luttinger liquid for the case of 
broken spin-symmetry. We obtain a significant homogeneous broadening of the ESR line-width with increasing 
temperature, which explains the unobservability of ESR in single-wall carbon nanotubes and puts severe constraints 
on the usability of SWCNTs for spintronics.

\begin{acknowledgments}
The authors acknowledge useful discussions with L. Forr\'{o} and an illuminating exchange of e-mails with A. De Martino.
Supported by the Hungarian State Grants (OTKA) F61733, K72613, NK60984, F68852, and K60576. VZ acknowledges the Bolyai 
programme of the HAS for support.
\end{acknowledgments}

\bibliography{luttref2}
\bibliographystyle{apsrev}

\end{document}